\documentstyle[preprint,pre,aps]{revtex}

\begin{document}

\tighten
\preprint{TITCMT-98-4}
\draft

\title{Quantum Annealing in the Transverse Ising Model}
\author{Tadashi Kadowaki and Hidetoshi Nishimori}
\address{Department of Physics, Tokyo Institute of Technology, Oh-okayama,
Megro-ku, Tokyo 152-8551, Japan}

\date{\today}
\maketitle

\begin{abstract}
We introduce quantum fluctuations into the simulated annealing process of
optimization problems, aiming at faster convergence to the optimal state.
Quantum fluctuations cause transitions between states
and thus play the same role as thermal fluctuations
in the conventional approach.
The idea is tested by the transverse Ising model,
in which the transverse field is a function of time similar to
the temperature in the conventional method.
The goal is to find the ground state of the diagonal part of
the Hamiltonian with high accuracy as quickly as possible.
We have solved the time-dependent Schr\"odinger equation numerically
for small size systems with various exchange interactions.
Comparison with the results of the corresponding classical
(thermal) method reveals
that the quantum annealing leads to the ground state with much larger
probability in almost all cases
if we use the same annealing schedule.
\end{abstract}


\narrowtext

\section{Introduction}
\label{sec:1}

The technique of simulated annealing (SA) was first proposed
by Kirkpatrick {\it et al.} \cite{Kirk}
as a general method to solve optimization problems.
The idea is to use thermal fluctuations to allow the system
to escape from local minima of the cost function
so that the system reaches the global minimum
under an appropriate annealing schedule
(the rate of decrease of temperature).
If the temperature is decreased too quickly,
the system may become trapped in a local minimum.
Too slow annealing, on the other hand, is practically
useless although such a process would certainly bring
the system to the global minimum.
Geman and Geman proved a theorem on the annealing schedule
for a generic problem of combinatorial optimization \cite{Geman}.
They showed that any system reaches the
global minimum of the cost function asymptotically
if the temperature is decreased as
$T= c/\log t$ or slower,
where $c$ is a constant determined by the system size and
other structures of the cost function.
This bound on the annealing schedule may be the optimal
one under generic conditions
although faster decrease of the temperature often
gives satisfactory results in practical applications
for many systems.

Thermal fluctuations
were introduced in the optimization problem so that
transitions between states take place in the process
of search for the global minimum among many states.
Thus there seems to be no reasons to avoid use of
other mechanisms for state transitions if
these mechanisms may lead to better convergence properties.
One such possibility is the generalized transition probability
of Tsallis \cite{Tsallis}, which is a generalization of the
conventional Boltzmann-type transition probability
appearing in the master equation and thus used
in Monte Carlo simulations
(see also \cite{Nishi}).
In the present paper we seek another possibility of making
use of quantum tunneling processes for state transitions,
which we call quantum annealing (QA).
In particular we would like to learn how effectively quantum
tunneling processes possibly lead to the global minimum
in comparison to temperature-driven processes used in the
conventional method of SA.

In the virtual absence of previous studies along such
a line of consideration,
it seems better to focus our attention on
a specific model system, rather than to develop a general
argument, to gain an insight into the role of quantum
fluctuations in the situation of optimization problem.
Quantum effects have been found to play a very similar role
to thermal fluctuations in the Hopfield model in a transverse
field in thermal equilibrium \cite{NN}.
This observation motivates us to investigate dynamical
properties of the Ising model
under quantum fluctuations
in the form of a transverse field.
We therefore discuss in this paper the transverse
Ising model with a variety of exchange interactions.
The transverse field controls the rate of transition
between states and thus plays the same role as
the temperature does in SA.
We assume that the system has no thermal fluctuations
in the QA context and the term ``ground state''
refers to the lowest-energy state of the Hamiltonian
without the transverse field term.

Static properties of the transverse Ising model have been
investigated quite extensively for many years \cite{Chak}.
There have however been very few studies on
the dynamical behavior of the Ising model with a transverse field.
We refer to the work by Sato {\it et al.} who carried out
quantum Monte Carlo simulations of the two dimensional gaussian spin glass
model in an infinitesimal transverse field, showing a reasonably fast approach
to the ground state \cite{Sato}.

We present here a new point of view by comparing the
efficiency of QA directly with that of classical SA in reaching
the ground state.
We solve the time-dependent Schr\"odinger equation and 
the classical master equation numerically for small-size systems
with the same exchange interactions under the same annealing schedules.
Calculations of probabilities that the system is in the ground
state at each time for both classical and quantum cases
give important implications on the relative efficiency of the
two approaches.

In the next section we explain the model and define the measure
of closeness of the system in QA to the desired ground state.
This measure is compared with the corresponding classical
probability that the system is in the ground state, the definition
of which is also given there.
In Sec. \ref{sec:3} numerical results for QA and SA are shown for various
types of annealing schedules and interactions.
The data suggest that QA generally gives a larger probability
to lead to the ground state than SA under the same
conditions on the annealing schedule and interactions.
Section \ref{sec:4} deals with the analytical solutions for the one-spin case
which turns out to be quite non-trivial.
Explicit solutions yield very useful information to clarify several
subtle aspects of the problem.
The final section is devoted to summary and discussions.

\section{Transverse Ising Model}
\label{sec:2}

Let us consider the following Ising model with longitudinal
and transverse fields:
\begin{eqnarray}
{\cal H}(t) & = &
 -\sum_{ij} J_{ij}\sigma^z_i\sigma^z_j -h\sum_i\sigma^z_i
 -\Gamma(t)\sum_i\sigma^x_i 
  \label{Hamiltonian}\\
 & \equiv & {\cal H}_0 - \Gamma(t)\sum_i\sigma^x_i,
\end{eqnarray}
where the type of interactions will be specified later.
The term of longitudinal field was introduced to remove
the trivial degeneracy in the exchange interaction term
coming from the overall up-down symmetry that effectively
reduces the available phase space by half.
The $\Gamma (t)$-term causes quantum tunneling
between various classical states
(the eigenstates of the classical part ${\cal H}_0$).
By decreasing the amplitude $\Gamma(t)$ of the transverse field
from a very large value to zero, we hopefully drive the system
into the optimal state, the ground state of ${\cal H}_0$.

The natural dynamics of the present system is provided by
the Schr\"odinger equation
\begin{equation}
i \frac{\partial \left|\psi(t)\right\rangle}{\partial t} =
 {\cal H}(t) \left|\psi(t)\right\rangle.
   \label{Schroedinger}
\end{equation}
We solve this time-dependent Schr\"odinger equation numerically
for small-size systems.
The representation to diagonalize ${\cal H}_0$ (the $z$-representation)
will be used throughout the paper.
The corresponding classical SA process is described by
the master equation
\begin{equation}
\frac{d P_i(t)}{d t} = \sum_j {\cal L}_{ij} P_j(t),
 \label{master}
\end{equation}
where $P_i(t)$ represents the probability that the system is
in the $i$th state.
We consider single-spin flip processes with 
the transition matrix elements given as
\widetext
\begin{equation}
{\cal L}_{ij} = \left\{
\begin{array}{ll}
 \left\{ 1 + \exp((E_i-E_j)/T(t)) \right\}^{-1} &
   (\mbox{single-spin difference}) \\
 - \sum_{k\neq i} {\cal L}_{ki} & (i=j)\\
 0 & (\mbox{otherwise})
\end{array}
\right. .
 \label{rate}
\end{equation}
\narrowtext

In SA, the temperature
$T(t)$ is first set to a very large value and then is gradually
decreased to zero.
The corresponding process in QA should be to change $\Gamma (t)$
from a very large value to zero.
The reason is that the high-temperature state in SA
is a mixture of all possible states with almost equal probabilities,
and the corresponding state in QA is the linear combination of
all states with equal amplitude in the $z$-representation,
which is the lowest eigenstate of the Hamiltonian (\ref{Hamiltonian})
for very large $\Gamma$.
The low-temperature state after a successful SA is the ground
state of ${\cal H}_0$, which should also be the eigenstate
of ${\cal H}(t)$ as $\Gamma (t)$ is reduced to zero sufficiently
slowly in QA.
Another justification of identification of $\Gamma$ and $T$ comes
from the fact that the $T=0$ phase diagram of the Hopfield model
in a transverse field has almost the same structure as the
equilibrium phase diagram of the conventional Hopfield model
at finite temperature if we identify the temperature axis of the
latter phase diagram with the $\Gamma$ axis in the former \cite{NN}.
We therefore change $\Gamma (t)$ in QA and $T(t)$ in SA from
infinity to zero with the same functional forms
$\Gamma (t)=T(t)=c/t, c/\sqrt{t}, c/\log (t+1)$
($t: 0 \to \infty$) or $-ct$ ($t: -\infty \to 0$).
The reason for choosing these functional forms are that either
they allow for analytical solutions in the single-spin case
as shown in Sec. \ref{sec:4} or for comparison with the Geman-Geman
bound mentioned in Sec. \ref{sec:1}.

To compare the performance of the two methods QA and  SA,
we calculate the probabilities
$P_{\text{QA}}(t) = 
\left| \langle g |\psi(t) \rangle \right|^2$
for QA and $P_{\text{SA}}(t) = P_g(t)$ for SA, 
where $P_g(t)$ is the probability to find the system
in the ground state at time $t$ in SA
and $| g \rangle$ is the ground-state
wave function of ${\cal H}_0$.
Note that we treat only small-size systems (the number of
spins $N=8$) and thus the ground state can be picked
out explicitly.
In the ideal situation $P_{\text{QA}}(t)$ and
$P_{\text{SA}}(t)$ will be very small initially
and increase towards 1 as $t\to \infty$.

It is useful to introduce another set of quantities
$P_{\text{SA}}^{\text{st}}(T)$
and $P_{\text{QA}}^{\text{st}}(\Gamma)$.
The former is the Boltzmann factor
of the ground state of ${\cal H}_0$
at temperature $T$ while the latter is defined
as $\left| \langle g |\psi_\Gamma \rangle \right|^2$,
where the wave function $\psi_\Gamma$ is the lowest-energy
stationary state
of the full Hamiltonian (\ref{Hamiltonian})
for a given fixed value of $\Gamma$.
In the quasi-static limit, the system follows equilibrium
in SA and thus
$P_{\text{SA}}(t)
=P_{\text{SA}}^{\text{st}}(T(t))$.
Correspondingly for QA, 
$P_{\text{QA}}(t)=
P_{\text{QA}}^{\text{st}}(\Gamma (t))$
when $\Gamma (t)$ changes sufficiently slowly.
Thus the differences between both sides of these two
equations give measures how closely the system follows
quasi-static states during dynamical process of annealing.

\section{Numerical results}
\label{sec:3}

We now present numerical results on
$P_{\text{SA}}$ and $P_{\text{QA}}$
for various types of exchange interactions and transverse fields.
All calculations were performed with a constant longitudinal field $h=0.1$
to remove trivial degeneracy.

\subsection{Ferromagnetic Model}

Let us first discuss the ferromagnetic Ising model
with $J=\mbox{const}$ for all pairs of spins.
Figure \ref{fig:1} shows the overlaps for the case
of $\Gamma (t)=T(t)=3/\log(t+1)$.
It is seen that both  QA and SA follow stationary
(equilibrium) states during dynamical processes
rather accurately.
In SA the theorem of Geman and Geman \cite{Geman} guarantees that
the annealing schedule $T(t)=c/\log (1+t)$ assures convergence
to the ground state ($P_{\text{SA}} \to 1$
in our notation) if $c$ is adjusted appropriately.
Our choice $c=3$ is somewhat arbitrary but the tendency
is clear for $P_{\text{SA}} \to 1$ as $t\to\infty$,
which is also clear from approximate satisfaction
of the quasi-equilibrium condition
$P_{\text{SA}}(t)
=P_{\text{SA}}^{\text{st}}(T(t))$.
Although there are no mathematically rigorous arguments for
QA corresponding to the Geman-Geman bound,
the numerical data indicate convergence to the ground state
under the annealing schedule $\Gamma (t)=3/\log(t+1)$
at least for the ferromagnetic system.
It should be remembered that the unit of time is arbitrary
since we have set $\hbar =1$ in the Schr\"odinger equation
(\ref{Schroedinger}) and the unit of time $\tau =1$
in the master equation (\ref{master}).
Thus the fact that the curves for QA in Fig. \ref{fig:1}
lie below those for SA at any given time does not
have any positive significance.

If we decrease the transverse field and the temperature faster,
$\Gamma (t)=T(t)=3/\sqrt {t}$,
there appears a qualitative difference between QA and SA as shown
in Fig. \ref{fig:2}.
The quantum method clearly gives better convergence to the ground state
while the classical counterpart gets stuck in a local minimum
with a non-negligible probability.
To see the rate of approach of $P_{\text{QA}}$
to 1, we have plotted $1-P_{\text{QA}}$ in a
log-log scale in Fig. \ref{fig:3}.
It is seen that $1-P_{\text{QA}}$ behaves
as ${\rm const}/t$ in the time region between 100 and 1000.

By a still faster annealing schedule $\Gamma (t)=T(t)=3/t$,
the system becomes trapped in intermediate states
both in QA and SA as seen in Fig. \ref{fig:4}.

\subsection{Frustrated Model}
\label{FrustModel}

We next analyze the interesting case of a frustrated system
shown in Fig. \ref{fig:5}.
The full lines indicate ferromagnetic interactions while
the broken line is for an antiferromagnetic interaction
with the same absolute value as the ferromagnetic ones.
If the temperature is very high in the classical case,
the spins 4 and 5 are changing their states very rapidly
and hence the effective interaction between spins 3 and 6
via spins 4 and 5 will be negligibly small.
Thus the direct antiferromagnetic interactions between spins
3 and 6 is expected to dominate the correlation of these
spins, which is clearly observed in Fig. \ref{fig:6}
as the negative value of the thermodynamic correlation function
$\langle \sigma_3^z \sigma_6^z\rangle_c$
in the high-temperature side.
At low temperatures, on the other hand, the spins 4 and 5 tend
to be fixed in some definite direction and consequently
the effective ferromagnetic interactions between spins 3 and 6
are roughly twice as large as the direct antiferromagnetic
interaction.
This argument is justified by the positive value of the correlation
function at low temperatures in Fig. \ref{fig:6}.
Therefore the spins 3 and 6 must change their relative orientation
at some intermediate temperature.
This means that the free-energy landscape goes under
significant restructuring as the temperature is decreased
and therefore the annealing process should be performed
with sufficient care.

If the transverse field in QA plays a similar role to the
temperature in SA, we expect similar dependence of the
correlation function $\langle \sigma_3^z \sigma_6^z\rangle_q$
on the transverse field $\Gamma$.
Here the expectation value is evaluated by the stationary
eigenfunction of the full Hamiltonian (\ref{Hamiltonian})
with the lowest eigenvalue at a given $\Gamma$.
The broken curve in Fig. \ref{fig:6} clearly supports
this idea.
We therefore expect that the frustrated system of Fig. \ref{fig:5}
is a good test ground for comparison of QA and SA in the situation
with a significant change of spin configurations in the dynamical
process of annealing.

The results are shown in Fig. \ref{fig:7} for the annealing
schedule $\Gamma (t)=T(t)=3/\sqrt{t}$.
The time scale is normalized so that both classical and quantum
correlation functions vanish at $t=1$ (see Fig. \ref{fig:6}).
The tendency is clear that QA is better suited for
ground-state search in the present system.

\subsection{Random Interaction Model}

The third and final example is the
Sherrington-Kirkpatrick (SK) model of spin glasses \cite{SK}.
Interactions exist between all pairs of spins and are chosen
from a Gaussian distribution with vanishing mean and
variance $1/N$ ($N=8$ in our case).
Figure \ref{fig:8} shows a typical result on the time
evolution of the probabilities under the annealing schedule
$\Gamma(t)=T(t)=3/\sqrt{t}$.
We have checked several realizations of exchange interactions
under the same distribution function and have found that
the results are qualitatively the same.
Figure \ref{fig:8} again suggests that QA is better suited
than SA for the present optimization problem.

\section{Solution of the single-spin problem}
\label{sec:4}

It is possible to solve the time-dependent Schr\"odinger equation
explicitly when the problem involves only a single spin and
the functional form of the transverse field is
$\Gamma (t)=-ct, c/t$ or $c/\sqrt{t}$.
We note that the single-spin problem is trivial
in SA because there are only two states involved (up and down)
and thus there are no local minima.
This does not mean that the same single-spin problem is also
trivial in the quantum mechanical version.
In QA with a single spin, the transition between the two states
is caused by a finite transverse field.
The system goes through tunneling processes
to reach the other state, and an approximate annealing schedule is
essential to reach the ground state.
On the other hand, in SA, the transition from the higher state
to the lower state takes place even at $T=0$
and thus the system always reaches the ground state.

Let us first discuss the case of $\Gamma (t)=-ct$ with
$t$ changing from $-\infty$ to 0.
This is the well-known Landau-Zener model and the explicit
solution of the time-dependent Schr\"odinger equation
is available in the literature \cite{Miya1,Zener,Miya2,Miya3,Raedt}.
With the notation $a(t)=\langle +|\psi (t)\rangle$ and
$b(t)=\langle -|\psi (t)\rangle$ and the initial condition
$a(-\infty)=b(-\infty)=1/\sqrt{2}$ (the lowest eigenstate),
the solution for $b(t)$ is found to be (see Appendix)
\begin{eqnarray}
\label{eq:lin}
\nonumber
b(t) & = & \frac{h e^{-\pi h^2/8c}}{2\sqrt{c}}
           \biggl\{-\frac{2ct+h}{h}D_{-\lambda-1}(-iz) \\
 & &       -\frac{ih^2+2c}{\sqrt{2c}h}e^{3/4 \pi i}
            D_{-\lambda-2}(-iz)\biggr\},
\end{eqnarray}
where $D_{-\lambda -1}, D_{-\lambda -2}$ represent
the parabolic cylinder function (or Weber function) and
$z$ and $\lambda$ are given as
\begin{eqnarray}
 z & = & \sqrt{2c}e^{-\pi i/4}t, \\
 \lambda & = & \frac{ih^2}{2c}.
\end{eqnarray}
The final value of $b(t)$ at $t=0$ is
\begin{eqnarray}
\nonumber
b(0)&=& -\frac{h\sqrt{\pi}2^{-ih^2/4c}e^{-\pi h^2/8c}}{2\sqrt{2c}} \\
 & & \times \biggl\{ \frac{1}{\Gamma(1+ih^2/4c)}
             +\frac{\sqrt{c}e^{3/4\pi i}(1+ih^2/2c)}
            {h\Gamma(3/2+ih^2/4c)}\biggr\}.
\end{eqnarray}
The probability to find the system in the ground state at $t=0$ is,
when $h^2/c \gg 1$,
\begin{equation}
P_{\text{QA}}(0) = |a(0)|^2 =  1-|b(0)|^2 \sim 1-\frac{9c^2}{16h^4}.
\end{equation}
Thus the probability $P_{\text{QA}}(t)$ does not approach
1 for finite $c$.

We next present the solution for $\Gamma (t)=c/t$
with $t$ changing from 0 to $\infty$ under the initial
condition $a=b=1/\sqrt{2}$ (see Appendix):
\begin{equation}
b(t) = \frac{1}{\sqrt{2}} e^{iht} t^{ic} F(1+ic,1+2ic;-2iht),
\end{equation}
where $F$ is the confluent hypergeometric function.
The asymptotic form of $b(t)$ as $t\to\infty$ is
\begin{eqnarray}
\nonumber
b(t) & \sim & \frac{\sqrt{2} (2h)^{-ic} \Gamma(2ic)}{\Gamma(ic)} \\
 & & \times \left\{ e^{-iht-\pi c/2} + c e^{iht+\pi c/2} (2ht)^{-1} \right\}.
\end{eqnarray}
The probability to find the system in the target ground state
behaves asymptotically as
\begin{eqnarray}
P_{\text{QA}}(t) & = & |a(t)|^2 \\
 & = & 1-|b(t)|^2 \\
\nonumber
 & \sim & 1- \frac{\sinh(\pi c)}{\sinh(2\pi c)} \\
 & &       \times \left\{e^{-\pi c} + \frac{c \cos(2ht)}{ht}
                + \frac{c^2 e^{\pi c}}{4h^2 t^2} \right\} \\
  & \sim & 1-e^{-2\pi c},
\end{eqnarray}
the last approximation being valid for $c \gg 1$ after
$t\to\infty$.
The system does not reach the ground state as $t\to\infty$ as
long as $c$ is finite.
Larger $c$ gives more accurate approach to the ground state,
which is reasonable because it takes longer time to reach
a given value of $\Gamma (=c/t)$ for larger $c$, implying
slower annealing.

The final example of the solvable model concerns the annealing
schedule $\Gamma (t)=c/\sqrt{t}$.
The solution for $b(t)$ is derived in Appendix under the
initial condition $a=b=1/\sqrt{2}$ as
\widetext
\begin{equation}
b(t) = \frac{1}{\sqrt{2}} e^{iht} 
   F\left(\frac{1}{2}-i\gamma,\frac{1}{2};-2iht \right)
     + \frac{c}{\sqrt{h}} e^{(3/4)\pi i} e^{iht} (-2iht)^{1/2}
                   F\left( 1-i\gamma,\frac{3}{2};-2iht \right),
\end{equation}
where $\gamma = c^2/2h$.
The large-$t$ behavior is found to be
%
\begin{eqnarray}
\nonumber
b(t) & \sim & \sqrt{\pi} e^{-\pi c^2/4h} \Biggl[
        e^{-iht} (2ht)^{-i\gamma} \left\{
        \frac{1}{\sqrt{2}\Gamma(\frac{1}{2}-i\gamma)}
        + \frac{\sqrt{h} e^{(5/4)\pi i}}{c\Gamma(-i\gamma)} \right\} \\
\label{asym}
 & &  + e^{iht} (2ht)^{-1/2+i\gamma} \left\{
        \frac{e^{-(1/4)\pi i}}{\sqrt{2}\Gamma(i\gamma)}
        + \frac{c}{2\sqrt{h}\Gamma(\frac{1}{2}+i\gamma)} \right\} \Biggr],
\end{eqnarray}
\narrowtext
\noindent
and the probability $P_{\text{QA}}(\infty)$ for $c^2/h\gg 1$
is obtained as
\begin{equation}
P_{\text{QA}}(\infty) = 1-|b(\infty)|^2 \sim 1-\frac{h^2}{64c^4}.
\end{equation}
This equation indicates that the single-spin system does not reach
the ground state under the present annealing schedule
$\Gamma (t)=c/\sqrt{t}$ for which the numerical data in the
previous section suggested an accurate approach.
We therefore conclude that the asymptotic value of
$P_{\text{QA}}(t)$ in the previous section may not
be exactly equal to 1 for $\Gamma(t)=3/\sqrt{t}$ although
it is very close to 1.

The annealing schedule $\Gamma (t)=c/\sqrt{t}$ has a feature
which distinguishes this function from the other ones
$-ct$ and $c/t$.
As we saw in the previous discussion, the final asymptotic value
of $P_{\text{QA}}(t)$ is not 1 if the initial
condition corresponds to the ground state for $\Gamma\to\infty$,
$a=b=1/\sqrt{2}$.
However, as shown in Appendix, by an appropriate choice of the initial
condition, it is possible to drive the system to the
ground state if $\Gamma (t)=c/\sqrt{t}$.
This is not possible for any initial conditions in the case
of $\Gamma (t)=-ct$ or $c/t$.

\section{Summary and Discussions}
\label{sec:5}

We have proposed the idea of quantum annealing (QA) in which
quantum tunneling effects cause transitions between states
in optimization problem, in contrast to the usual
thermal transitions in simulated annealing (SA).
The idea was tested in the transverse Ising model obeying
the time-dependent Schr\"odinger equation.
The transverse field term was controlled so that the system
approaches the ground state.
The numerical results on the probability to find the system
in the ground state were compared with the corresponding
probability derived from the numerical solution of the 
master equation representing the SA processes.
We have found that QA shows convergence to the optimal (ground)
state with larger probability than SA in all cases if the
same annealing schedule is used.
The system approaches the ground state rather accurately
in QA for the annealing schedule $\Gamma =c/\sqrt{t}$ but
not for faster decrease of the transverse field.

We have also solved the single-spin model exactly for QA
in the cases of $\Gamma (t)=-ct, c/t$ and $c/\sqrt{t}$.
The results showed that the ground state is not reached
perfectly for all these annealing schedules.
Therefore the asymptotic values of $P_{\text{QA}}(t)$
in numerical calculations are probably not exactly 1
although they seem to be quite close to the optimal value 1.

The rate of approach to the asymptotic value close to 1,
$1-P_{\text{QA}}(t)$, was found to
be proportional to $1/t$ in Fig. \ref{fig:3} for the ferromagnetic
model.
On the other hand, the single-spin solution shows the existence
of a term proportional to $1/\sqrt{t}$, see Eq. (\ref{asym}).
Probably the coefficient of the $1/\sqrt{t}$-term is very small
in the situation of Fig. \ref{fig:3}
and the next-order contribution dominates in the time region
shown in Fig. \ref{fig:3}.

A simple argument using perturbation theory yields useful
information about the asymptotic form of the probability
function if we assume that the system follows quasi-static
states during dynamical processes.
The probability to find the system in the ground state is expressed
using the perturbation in terms of $\Gamma (\ll 1)$ as
\begin{equation}
 P_{\text{QA}}(\Gamma)
  \sim  1-\Gamma^2\sum_{i\neq 0}\frac{1}{(E_0^{(0)}-E_i^{(0)})^2},
\end{equation}
where $E_i^{(0)}$ is the energy of the $i$th state of
the non-perturbed (classical) system
and $E_0^{(0)}$ is the ground-state energy.
If we set $\Gamma=c/\sqrt{t}$, we have
\begin{equation}
 P_{\text{QA}}(\Gamma) \sim
   1-\frac{1}{t}\sum_{i\neq 0}\left(\frac{c}{E_0^{(0)}-E_i^{(0)}}\right)^2.
\end{equation}
Thus the approach to the asymptotic value is proportional to $1/t$
as long as the system stays in quasi-static states.
The corresponding probability for SA is
\begin{equation}
 P_{\text{SA}}(T) \sim \frac{e^{-E_0/T}}{\sum_i e^{-E_i/T}}
  \sim 1-\sum_{i\neq 0} e^{-(E_i-E_0)/T},
\end{equation}
which shows absence of universal ($1/t$-like) dependence on time.

The present method of QA bears some similarity to the approach by
the generalized transition probability in which
the dynamics is described by the master equation but the transition
probability has power-law dependence on the temperature in contrast
to the usual exponential form of the Boltzmann factor
\cite{Tsallis}.
This power-law dependence on the temperature allows the system
to search for a wider region in the phase space
because of larger probabilities of transition to higher-energy
states at a given $T(t)$, which may be the reason of faster
convergence to the optimal states \cite{Tsallis,Nishi}.
The transverse field term $\Gamma$ in our QA represents the rate of
transition between states which is larger than the transition rate
in SA (see (\ref{rate})) at a given small value of
the control parameter $\Gamma (t)=T(t)$.
This larger transition probability may lead to a more active search
in wider regions of the phase space, leading to better
convergence similarly to the case of the generalized transition
probability.

We have solved the Schr\"odinger equation and the master equation directly
by numerical methods for the purpose of comparison of QA and SA.
This method faces difficulties for larger $N$ because the number
of states increases exponentially as $2^N$.
The classical SA solves this problem by exploiting stochastic processes,
Monte Carlo simulations, which have the computational complexity
growing as a power of $N$.
The corresponding reduction of the computational complexity is lacking
in QA, and it is an important future problem in practical implementation
of the idea of QA.
Another future problem is to devise implementations of QA in other
optimization problems such as the travelling salesman problem or
the graph bipartitioning for which there seems to be no direct
analogue of the transverse field to cause quantum transitions.

\acknowledgments

We acknowledge useful discussions
with Prof. S. Miyashita who suggested the model and analysis 
in Sec. \ref{FrustModel}.
One of the authors (T.K.) is grateful for the financial support
of the Japan Society for the Promotion of Science
for Japanese Junior Scientists.

\appendix
\section*{Single-spin problem}

In this Appendix we explain some technical aspects to derive
the exact solution of the time-dependent Schr\"odinger equation
for the transverse Ising model with a single spin.
The three cases of
$\Gamma(t) = - c t, c / t$ and  $c / \sqrt{t}$
will  be discussed.

\subsection{Case of $\Gamma(t) = - c t$ (Landau-Zener model)
\protect\cite{Miya1,Zener,Miya2,Miya3,Raedt}}

Let us express the solution of the Schr\"odinger equation at time $t$
by the parameters $a=\langle +|\psi (t)\rangle$ and
$b=\langle -|\psi (t)\rangle$.
The Schr\"odinger equation (\ref{Schroedinger}) with
${\cal H}=-h\sigma^z-\Gamma \sigma^x$ is expressed as
a set of first order differential equations for $a$ and $b$.
It is convenient to change the variables as
\begin{equation}
 \tilde{a}=\frac{1}{\sqrt{2}}(a+b), \hspace{1em}
 \tilde{b}=\frac{1}{\sqrt{2}}(a-b),
\end{equation}
by which the Schr\"odinger equation is now
\begin{equation}
 \frac{d^2\tilde{b}(t)}{d t^2}+(-ic+h^2+c^2t^2)\tilde{b}(t)=0.
\end{equation}
By using the notation
\begin{eqnarray}
 z & = & \sqrt{2c}e^{-\pi i/4}t, \\
 \lambda & = & \frac{ih^2}{2c},
\end{eqnarray}
we find
\begin{equation}
 \frac{d^2\tilde{b}(t)}{d z^2}
  +(\lambda+\frac{1}{2}-\frac{1}{4}z^2)\tilde{b}(t)=0.
 \label{diff1}
\end{equation}

The initial state is specified as $a=b=1/\sqrt{2}$ or
$\tilde{b}=0$ as $t\to -\infty$.
The solution of (\ref{diff1}) satisfying this condition is the
parabolic cylinder function $D_{-\lambda-1}(-iz)$ \cite{specialf}.
Thus, we obtain the solution as
\begin{eqnarray}
 \tilde{a}(t) & = & \frac{1}{h}\left(-ct\tilde{b}(t)
                    -i\frac{d\tilde{b}(t)}{dt}\right), \\
 \tilde{b}(t) & = & C_1D_{-\lambda-1}(iz),
\end{eqnarray}
where $C_1$ is a constant.
To fix $C_1$, we use the condition
\begin{equation}
|\tilde{a}(-\infty)| = \frac{2C_1ce^{\pi h^2/8c}}{h\sqrt{2c}} = 1.
\end{equation}
Then we have
\begin{equation}
 C_1=\frac{h}{\sqrt{2c}}e^{-\pi h^2/8c}.
\end{equation}
The wave function of this system is given in Eq. (\ref{eq:lin}).

\subsection{Case of $\Gamma(t) = c/t$}

We next consider the case of $\Gamma(t) = c/t$.
By eliminating $a$ from the Schr\"odinger equation, we obtain
\begin{eqnarray}
\nonumber
\lefteqn{\frac{d^2 b(t)}{d t^2}
 - \frac{1}{\Gamma(t)}\frac{d\Gamma(t)}{dt}\frac{db(t)}{dt}} \\
\label{eq:sch_e}
 & & + \left(h^2+\Gamma^2(t)-\frac{ih}{\Gamma(t)}\frac{d\Gamma(t)}{dt}\right)
 b(t) = 0.
\end{eqnarray}
Substituting $\Gamma(t) = c/t$, we have
\begin{equation}
 \frac{d^2 b(t)}{d t^2}
 - \frac{1}{t}\frac{d b(t)}{d t}
 + \left( h^2 + \frac{i h}{t} + \frac{c^2}{t^2}\right) b(t) = 0.
\end{equation}
The solutions of this equation are expressed by the confluent $P$ function
\cite{specialf}
\begin{eqnarray}
\nonumber
\lefteqn{\tilde{P} \left\{
 \begin{array}{cccc}
  \infty & 0 & \\
  \overbrace{ \makebox[5ex]{$ih$} \makebox[3ex]{$1$} } & ic & t \\
  \makebox[5ex]{$-ih$} \makebox[3ex]{$0$} & -ic &
 \end{array}
 \right\}} \\
 & & = e^{iht} t^{ic} \tilde{P} \left\{
 \begin{array}{cccc}
  \infty & 0 & \\
  \overbrace{ \makebox[5ex]{$0$} \makebox[6ex]{$1+ic$} } & 0 & -2iht \\
  \makebox[5ex]{$1$} \makebox[6ex]{$ic$} & -2ic &
 \end{array}
 \right\},
\end{eqnarray}
the right-hand side of which has two independent expressions
in terms of the confluent hypergeometric function
\begin{eqnarray}
 f(t) & = & e^{iht} t^{ic} F(1+ic,1+2ic;-2iht), \\
\nonumber
 g(t) & = & e^{iht} t^{ic} (-2iht)^{-2ic} \\
      &   & \times F(1-ic,1-2ic;-2iht).
\end{eqnarray}
The general solution is $b(t) = C_1 f(t) + C_2 g(t)$.
Using the initial condition
\begin{eqnarray}
b(0) & = & C_1+C_2 = \frac{1}{\sqrt{2}}, \\
a(0) & = & C_1-C_2 = \frac{1}{\sqrt{2}},
\end{eqnarray}
we find
\begin{equation}
b(t) = \frac{1}{\sqrt{2}} e^{iht} t^{ic} F(1+ic,1+2ic;-2iht).
\end{equation}
The asymptotic forms of $b(t)$ and $|b(t)|^2$ are then given as
\begin{eqnarray}
\nonumber
b(t) & \sim & \frac{\sqrt{2} (2h)^{-ic} \Gamma(2ic)}{\Gamma(ic)} \\
 & &  \times \left\{e^{-iht-\pi c/2} + c e^{iht+\pi c/2} (2ht)^{-1} \right\},
\end{eqnarray}
\begin{equation}
|b(t)|^2 \sim \frac{\sinh(\pi c)}{\sinh(2\pi c)} \left\{
e^{-\pi c} + \frac{c \cos(2ht)}{ht} + \frac{c^2 e^{\pi c}}{4h^2 t^2} \right\}.
\end{equation}

\subsection{Case of $\Gamma(t) = c/\sqrt{t}$}

The final solvable model has $\Gamma(t) = c/\sqrt{t}$.
The Schr\"odinger equation (\ref{eq:sch_e}) is then expressed as
\begin{equation}
\frac{d^2 b(t)}{d t^2}
 - \frac{1}{2t}\frac{d b(t)}{d t}
 + \left( h^2 + \frac{2c^2+ih}{2t} \right) b(t) = 0.
 \label{diff2}
\end{equation}
The solution is the confluent $P$ function \cite{specialf}
\begin{eqnarray}
\nonumber
\lefteqn{\tilde{P} \left\{
 \begin{array}{cccc}
  \infty & 0 & \\
  \overbrace{\makebox[5ex]{$ih$}\makebox[9ex]{$\frac{1}{2}-i\gamma$}} & 0 & t\\
  \makebox[5ex]{$-ih$} \makebox[9ex]{$i\gamma$} & \frac{1}{2} &
 \end{array}
 \right\}} \\
 & & = e^{iht} \tilde{P} \left\{
 \begin{array}{cccc}
  \infty & 0 & \\
  \overbrace{\makebox[5ex]{$0$}\makebox[9ex]{$\frac{1}{2}-i\gamma$}} & 0 &
   -2iht \\
  \makebox[5ex]{$1$} \makebox[9ex]{$i\gamma$} & \frac{1}{2} &
 \end{array}
 \right\},
\end{eqnarray}
where $\gamma = c^2/2h$.
The two independent solutions are thus \cite{specialf}
\begin{eqnarray}
 f(t) & = & e^{iht} F(\frac{1}{2}-i\gamma,\frac{1}{2};-2iht), \\
 g(t) & = & e^{iht} (-2iht)^{1/2} F(1-i\gamma,\frac{3}{2};-2iht).
\end{eqnarray}
The general solution of (\ref{diff2}) is therefore the linear combination
of the above two functions
\begin{equation}
\label{eq:sol_lin}
 b(t) = C_1 f(t) + C_2 g(t).
\end{equation}
The constants $C_1$ and $C_2$ are fixed by the requirement
\begin{eqnarray}
 b(0) & = & C_1 = \frac{1}{\sqrt{2}}, \\
 a(0) & = & \frac{\sqrt{h}}{\sqrt{2}c}e^{(5/4)\pi i}C_2  = \frac{1}{\sqrt{2}}.
\end{eqnarray}
Substituting $C_1$ and $C_2$ into Eq. (\ref{eq:sol_lin}),
we find
\widetext
\begin{equation}
b(t) = \frac{1}{\sqrt{2}} e^{iht} 
    F\left(\frac{1}{2}-i\gamma,\frac{1}{2};-2iht \right)
     + \frac{c}{\sqrt{h}} e^{(3/4)\pi i} e^{iht} (-2iht)^{1/2}
                   F\left(1-i\gamma,\frac{3}{2};-2iht \right).
\end{equation}
The asymptotic form is
%
\begin{eqnarray}
\nonumber
b(t) & \sim & \sqrt{\pi} e^{-\pi c^2/4h} \Biggl[
           e^{-iht} (2ht)^{-i\gamma} \left\{
           \frac{1}{\sqrt{2}\Gamma(\frac{1}{2}-i\gamma)}
           + \frac{\sqrt{h} e^{(5/4)\pi i}}{c\Gamma(-i\gamma)} \right\} \\
\label{asym2}
 & &     + e^{iht} (2ht)^{-1/2+i\gamma} \left\{
           \frac{e^{-(1/4)\pi i}}{\sqrt{2}\Gamma(i\gamma)}
           + \frac{c}{2\sqrt{h}\Gamma(\frac{1}{2}+i\gamma)} \right\} \Biggr].
\end{eqnarray}
The probability $|b(\infty)|^2$ that the system remains in the excited state
can be calculated as the asymptotic
form of (\ref{asym2}) with the condition $c^2/h\gg 1$
\begin{eqnarray}
|b(\infty)|^2 & = & \frac{\pi e^{-\gamma \pi}}{2}
                \left|\frac{1}{\Gamma(\frac{1}{2}-i\gamma)}
             +\frac{\gamma^{-1/2} e^{(5/4)\pi i}}{\Gamma(-i\gamma)}\right|^2 \\
 & \sim & \frac{e^{-\gamma \pi}}{4}
                \left|e^{1/2}\left(\frac{1}{2}-i\gamma\right)^{i\gamma}
               +\gamma^{-1/2}e^{(5/4)\pi i}(-i\gamma)^{i\gamma+1/2}\right|^2 \\
 & \sim & \frac{e^{-\gamma\pi}}{4}\left|(-i\gamma)^{i\gamma}\frac{i}{8\gamma}
               \right|^2 = \frac{1}{256\gamma} = \frac{h^2}{64c^4}.
\end{eqnarray}
\narrowtext

\subsection{Dependence of the final value on the initial condition}

We show that we can choose the initial condition so that
the final state is the ground state when $\Gamma=c/\sqrt{t}$.
This is not possible for $\Gamma=-ct$ or $c/t$.
From Eq. (\ref{eq:sol_lin}), the asymptotic form of
the solution as $t\rightarrow\infty$ is
\begin{eqnarray}
\nonumber
b(t) & \sim & C_1 \frac{\sqrt{\pi}e^{-\pi c^2/4h-iht}(2ht)^{-ic^2/2h}}
           {\Gamma(1/2-ic^2/2h)} \\
     &   &+C_2 \frac{i\sqrt{\pi}he^{-\pi c^2/4h-iht}(2ht)^{-ic^2/2h}}
           {c^2\Gamma(-ic^2/2h)}.
\end{eqnarray}
The coefficients $C_1$ and $C_2$ are fixed under the conditions
$b(\infty)=0$ and $|a(0)|^2+|b(0)|^2=1$ as
\begin{eqnarray}
C_1 & = & \left\{1
         +\frac{\sinh(\pi c^2/h)}{2\sinh^2(\pi c^2/2h)}\right\}^{-1/2},\\
C_2 & = & \frac{ic^2\Gamma(-ic^2/2h)}{h\Gamma(1/2-ic^2/2h)}C_1.
\end{eqnarray}
This solution is not the ground state of the Hamiltonian ${\cal H}(0)$.

The reason why one cannot obtain such a solution for the other schedules
($\Gamma=c/t,-ct$) is the following:
The general solution for $\Gamma=c/t$ also has two coefficients,
and the initial state is represented as the linear combination
of two terms whose phases are indefinite:
\begin{eqnarray}
\label{eq:indet1}
a(0) & = & \left. C_1 t^{ic} \right|_{t\rightarrow 0} -
           \left. C_2 t^{-ic} \right|_{t\rightarrow 0},\\
\label{eq:indet2}
b(0) & = & \left. C_1 t^{ic} \right|_{t\rightarrow 0} +
           \left. C_2 t^{-ic} \right|_{t\rightarrow 0}.
\end{eqnarray}
The lowest-energy state at $t=0$ corresponds to $a(0)=b(0)=1/\sqrt{2}$
(times an arbitrary phase factor), which is realized by choosing $C_2=0$
in Eqs. (\ref{eq:indet1}) and (\ref{eq:indet2}).
The indefiniteness of $t^{ic}$ as $t\rightarrow 0$ is irrelevant
because this is only the overall phase factor.
Such a situation does not happen for other values of $a(0)$ and $b(0)$,
leading to a serious difficulty to determine the wave function at $t=0$.
Thus we cannot choose an initial condition other than $a(0)=b(0)=1/\sqrt{2}$.
A similar fact exists in the case of $\Gamma=-ct$.



\begin{figure}
\caption{Time dependence of the overlaps $P_{\text{SA}}(t)$,
$P_{\text{QA}}(t)$,
$P_{\text{SA}}^{\text{st}}(T(t))$
and $P_{\text{QA}}^{\text{st}}(\Gamma(t))$
of the ferromagnetic model with $\Gamma (t)=T(t)=3/\log(t+1)$.}
\label{fig:1}
\end{figure}

\begin{figure}
\caption{Time dependence of the overlaps
of the ferromagnetic model with $\Gamma (t)=T(t)=3/\sqrt{t}$.}
\label{fig:2}
\end{figure}

\begin{figure}
\caption{Time dependence of $1-P_{\text{QA}}(t)$
of the ferromagnetic model with $\Gamma (t)=3/\sqrt{t}$.
The dotted line represents $t^{-1}$ to guide the eye.
}
\label{fig:3}
\end{figure}

\begin{figure}
\caption{Time dependence of the overlaps
of the ferromagnetic model with $\Gamma (t)=T(t)=3/t$.}
\label{fig:4}
\end{figure}

\begin{figure}
\caption{The frustrated model where the
solid lines denote ferromagnetic interactions and the broken line
is for an antiferromagnetic interaction.
}
\label{fig:5}
\end{figure}

\begin{figure}
\caption{Correlation functions of spins 3 and 6 in Fig. \ref{fig:5}
for the classical and quantum cases.
In the classical model (full line) the correlation is shown as a function
of temperature while the quantum case (dotted line) is regarded as a function
of the transverse field.
}
\label{fig:6}
\end{figure}

\begin{figure}
\caption{Time dependence of the overlaps
of the frustrated model under
$\Gamma (t)=T(t)=3/\sqrt{t}$.
Here $\Gamma_{\text{c}}$ and $T_{\text{c}}$ are the points where the
correlation functions vanish in Fig. \ref{fig:6}.
}
\label{fig:7}
\end{figure}

\begin{figure}
\caption{Time dependence of the overlaps for
the SK model with $\Gamma (t)=T(t)=3/\sqrt{t}$.}
\label{fig:8}
\end{figure}

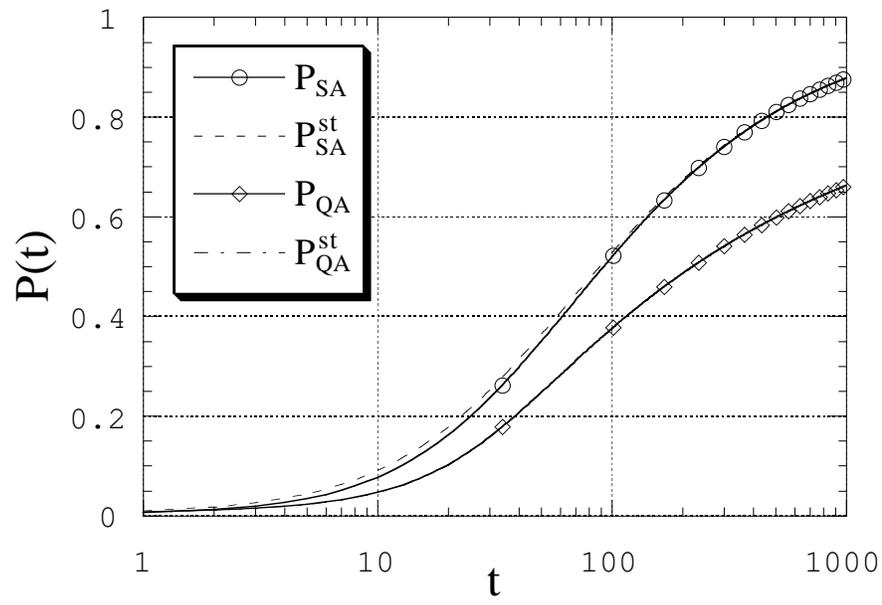
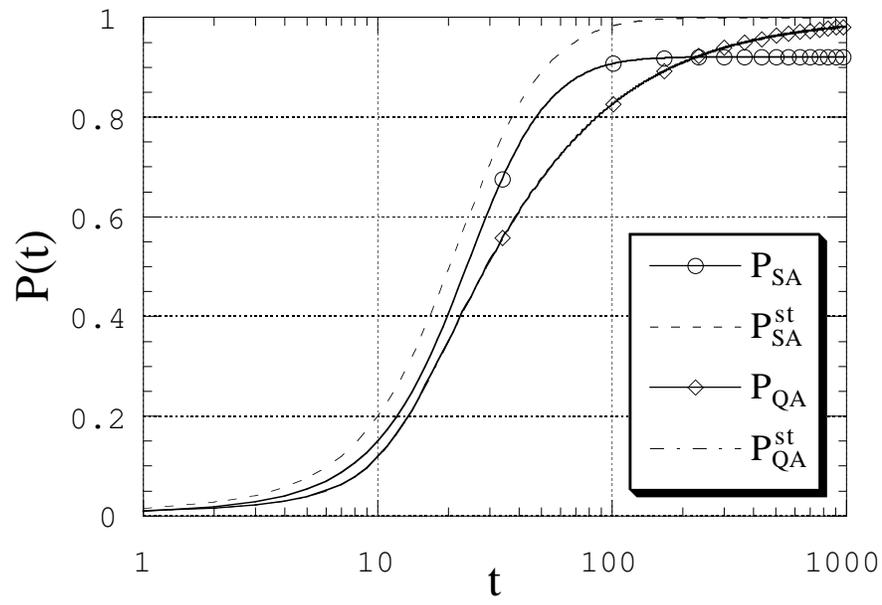
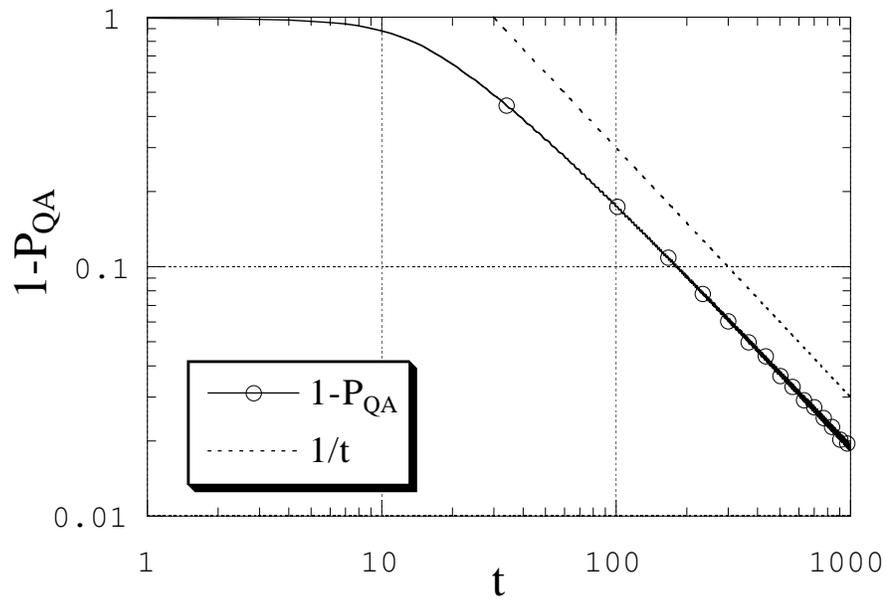
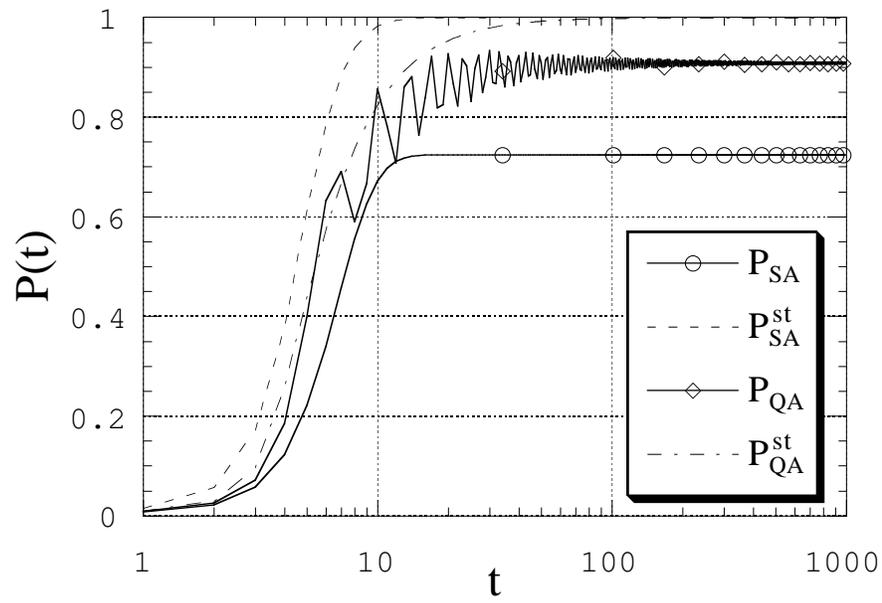
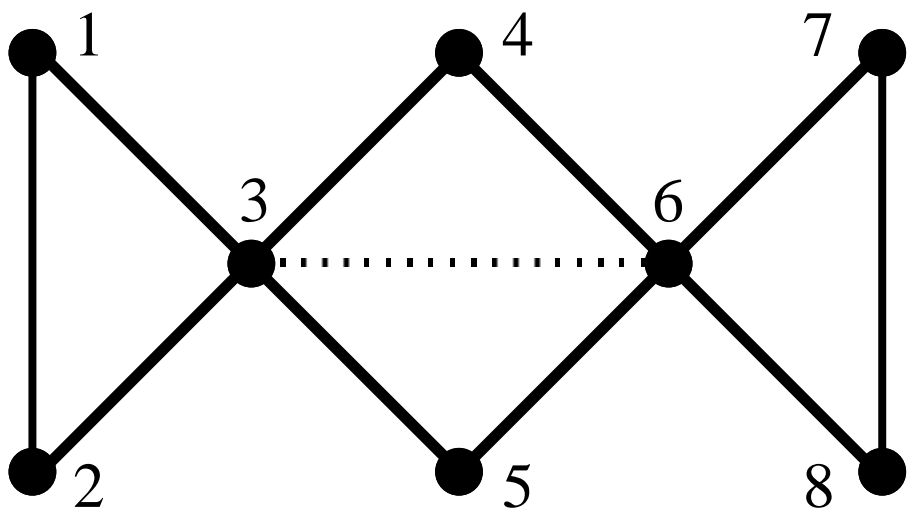
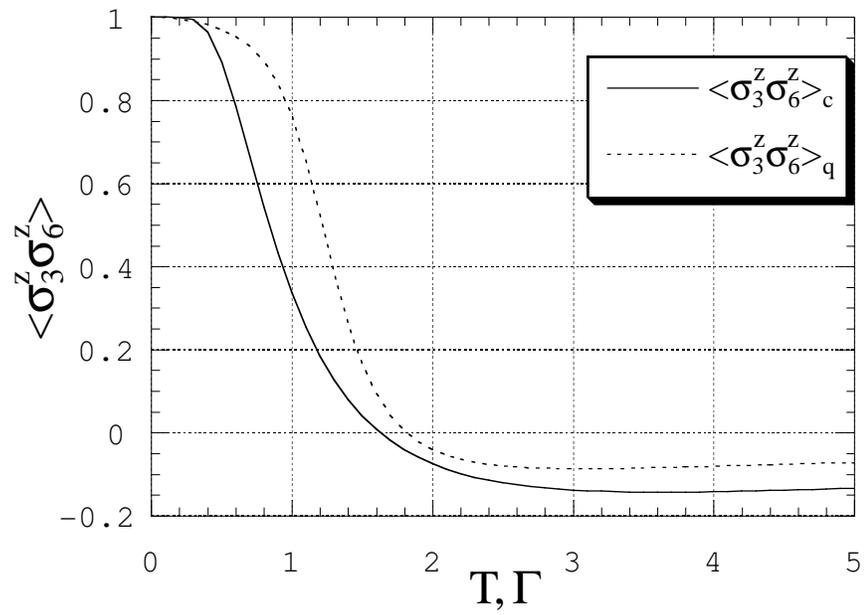
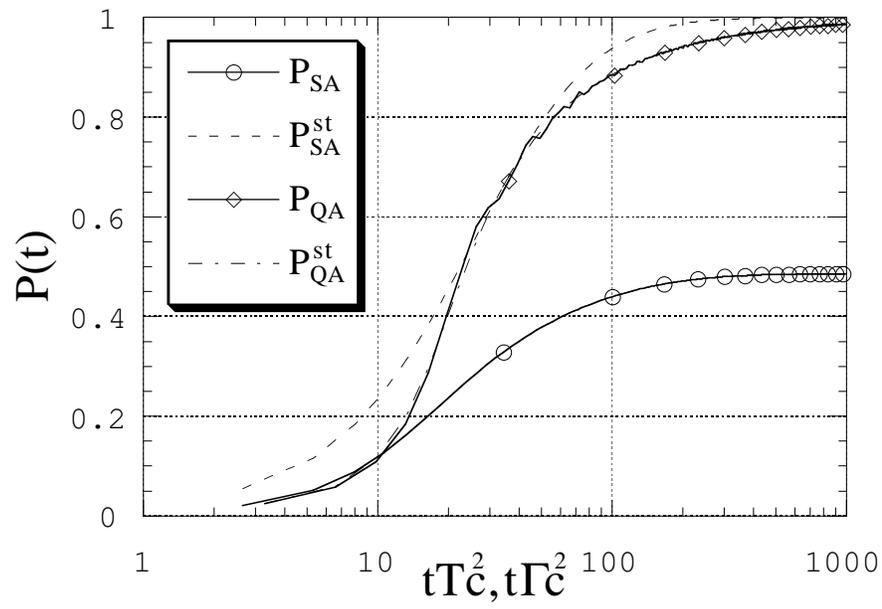
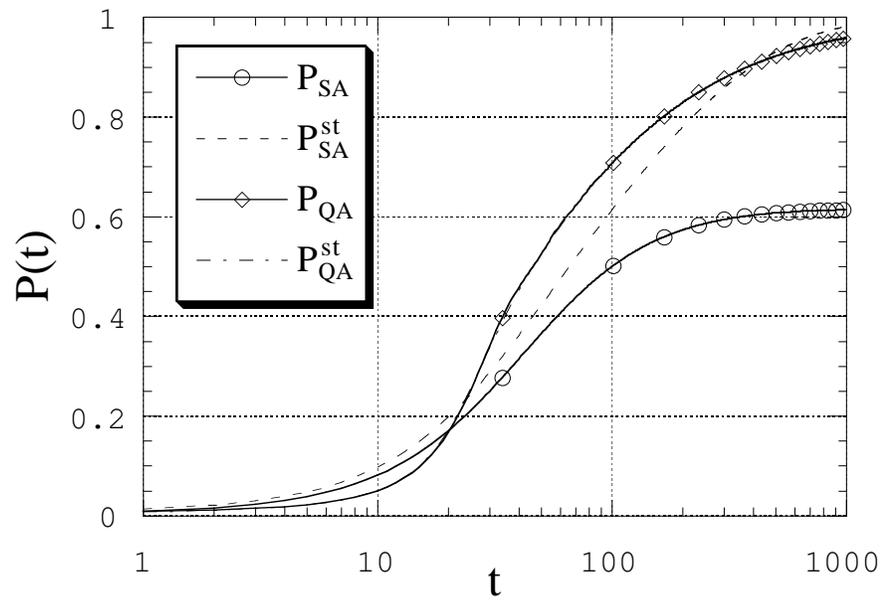
\end{document}